# The need to implement FAIR principles in biomolecular simulations

Amaro, R.[1]; Åqvist, J.[2]; Bahar, I.[3,4]; Battistini, F.[5]; Bellaiche, A.[6]; Beltran, D.[5]; Biggin, P.C.[7]; Bonomi, M.[8]; Bowman, G.R.[9]; Bryce, R.[10]; Bussi, G.[11]; Carloni, P.[12,13]; Case, D.[14]; Cavalli, A.[15,16]; Chang, C.A.[17]; Cheatham III, T.E.[18]; Cheung, M.S.[19,20]; Chipot, C.[21,22,23]; Chong, L.T.[24]; Choudhary, P.[6]; Cisneros, G.A.[25,26]; Clementi, C.[27]; Collepardo-Guevara, R.[28,29,30]; Coveney, P.[31,32]; Covino, R.[33,34]; Crawford, T.D.[35,36]; Dal Peraro, M.[37]; de Groot, B.[38]; Delemotte, L.[39]; De Vivo, M.[40]; Essex, J.[41]; Fraternali, F.[42]; Gao, J.[43]; Gelpí, J.L.[44,45]; Gervasio, F.L.[46,47,48,49]; Gonzalez-Nilo, F.D.[50]; Grubmüller, H.[51]; Guenza, M.G.[52]; Guzman, H.V.[53]; Harris, S.[54]; Head-Gordon, T.[55]; Hernandez, R.[56]; Hospital, A.[5,57]; Huang, N.[58]; Huang, X[59]; Hummer, G.[60,61]; Iglesias-Fernández, J.[62]; Jensen, J.H.[63]; Jha, S.[64]; Jiao, W.[65]; Jorgensen, W.L.[66]; Kamerlin, S.C.L.[67,68]; Khalid, S.[7] Laughton, C[69]; Levitt, M[70]; Limongelli, V.[71]; Lindahl, E.[72,39]; Lindorff-Larsen, K.[73]; Loverde, S.[74]; Lundborg, M.[39]; Luo, Y.L.[75]; Luque, F.J.[76,77]; Lynch, C.I.[7]; MacKerell, A.[78]; Magistrato, A.[79]; Marrink, S.J.[80]; Martin, H.[31]; McCammon, J.A.[81,82]; Merz, K.[83,84]; Moliner, V.[85]; Mulholland, A.[86]; Murad, S.[87]; Naganathan, A.N.[88]; Nangia, S.[89]; Noe, F.[90,91,92,93]; Noy, A.[94]; Oláh, J.[95]; O'Mara, M.[96]; Ondrechen, M.J.[97]; Onuchic, J.N.[98,92,99,100]; Onufriev, A.[101,102,103]; Osuna, S.[104,105]; Palermo G.[106,17]; Panchenko, A.R.[107,108,109]; Pantano, S.[110]; Parish, C[111]; Parrinello, M[112]; Perez, A.[113]; Perez-Acle, T.[114,115]; Perilla, J.R.[116,117]; Pettitt, B.M.[118]; Pietropaolo, A.[119]; Piquemal, J-P.[120]; Poma, A.[121]; Praprotnik, M.[122,123]; Ramos, M.J.[124]; Ren, P.[125]; Reuter, N.[126]; Roitberg, A.[113]; Rosta, E.[127]; Rovira, C.[128,105]; Roux, B.[129]; Röthlisberger, U.[130]; Sanbonmatsu, K.[131,132]; Schlick, T.[133]; Shaytan, A.K.[134]; Simmerling, C.[3,135,136]; Smith, J.C.[137,138]; Sugita, Y.[139,140,141]; Świderek, K.[85]; Taiji, M.[142]; Tao, P.[143]; Tikhonova, I.G.[144]; Tirado-Rives, J.[66]; Tuñón I.[145]; Van Der Kamp, M.W.[146]; Van der Spoel, D.[2]; Velankar, S.[6]; Voth, G.A.[147]; Wade, R.[148]; Warshel, A.[149]; Welborn, V.V.[150,36]; Wetmore, S.[151]; Wong, C.F.[152]; Yang, L-W.[153]; Zacharias, M.[154]; Orozco, M.[5,45]

1. Department of Molecular Biology, University of California San Diego, California, United States
2. Department of Cell and Molecular Biology, Uppsala University, Uppsala, Sweden
3. Laufer Center for Physical and Quantitative Biology, Renaissance School of Medicine, Stony Brook University, Stony Brook, New York, United States
4. Department of Biochemistry and Cell Biology, Renaissance School of Medicine, Stony Brook University, Stony Brook, New York, United States
5. Institute for Research in Biomedicine (IRB Barcelona), Barcelona, Spain
6. European Molecular Biology Laboratory, European Bioinformatics Institute, Hinxton, United Kingdom
7. Structural Bioinformatics and Computational Biochemistry, Department of Biochemistry, University of Oxford, Oxford, United Kingdom
8. Institut Pasteur, Université Paris Cité, CNRS UMR 3528, Computational Structural Biology Unit, Paris, France
9. Department of Biochemistry and Biophysics, University of Pennsylvania, Philadelphia, Pennsylvania, United States
10. Division of Pharmacy and Optometry, University of Manchester, Manchester, United Kingdom
11. Scuola Internazionale Superiore di Studi Avanzati-SISSA, Trieste, Italy
12. Computational Biomedicine, Institute of Advanced Simulations IAS-5/Institute for Neuroscience and Medicine INM-9, Forschungszentrum Jülich GmbH, Jülich, Germany
13. Department of Physics and Universitätsklinikum, RWTH Aachen University, Aachen, Germany
14. Department of Chemistry & Chemical Biology, Rutgers University, Piscataway, New Jersey, United States
15. Istituto Italiano di Tecnologia, Drug Discovery and Development, Bologna, Italy
16. Centre Européen de Calcul Atomique et Moléculaire (CECAM), Ecole Polytechnique Fédérale de Lausanne (EPFL), Lausanne, Switzerland
17. Department of Chemistry, University of California, Riverside, California, United States
18. Department of Medicinal Chemistry, College of Pharmacy, University of Utah, Salt Lake City, Utah, United States
19. Department of Physics, University of Washington, Seattle, Washington, United States
20. Pacific Northwest National Laboratory, Richland, Washington, United States
21. LIA CNRS-UIUC, UMR n°7019, Université de Lorraine, Vandœuvre-lès-Nancy, France
22. Department of Physics, University of Illinois at Urbana-Champaign, Urbana, Illinois, United States
23. Department of Biochemistry and Molecular Biology, The University of Chicago, Chicago, Illinois, United States
24. Department of Chemistry, University of Pittsburgh, Pittsburgh, Pennsylvania, USA
25. Department of Chemistry and Biochemistry, University of Texas at Dallas, Richardson, Texas, United States
26. Department of Physics, University of Texas at Dallas, Richardson, Texas, United States
27. Theoretical and Computational Biophysics, Department of Physics, Freie Universität Berlin, Berlin, Germany
28. Maxwell Centre, Cavendish Laboratory, Department of Physics, University of Cambridge, Cambridge, United Kingdom
29. Yusuf Hamied Department of Chemistry, University of Cambridge, Cambridge, United Kingdom
30. Department of Genetics, University of Cambridge, Cambridge, United Kingdom
31. Centre for Computational Science, University College London, London, United Kingdom
32. Institute for Informatics, University of Amsterdam, Amsterdam, Netherlands
33. Frankfurt Institute for Advanced Studies, Frankfurt am Main, Germany
34. Department of Biochemistry, University of Bayreuth, Bayreuth, Germany
35. Molecular Sciences Software Institute, Blacksburg, Virginia, United States
36. Department of Chemistry, Virginia Tech, Blacksburg, Virginia, United States
37. Institute of Bioengineering, School of Life Sciences, Ecole Polytechnique Fédérale de Lausanne (EPFL), Lausanne, Switzerland

38. Computational Biomolecular Dynamics Group, Department of Theoretical and Computational Biophysics, Max Planck Institute for Multidisciplinary Sciences, Göttingen, Germany
39. Department of Applied Physics, Science for Life Laboratory, KTH Royal Institute of Technology, Solna, Sweden
40. Laboratory of Molecular Modelling & Drug Discovery, Istituto Italiano di Tecnologia, Genoa, Italy
41. School of Chemistry, University of Southampton, Southampton, United Kingdom
42. Institute of Structural and Molecular Biology, University College London, London, United Kingdom
43. Department of Chemistry, University of Minnesota, Minneapolis, Minnesota, United States
44. Barcelona Supercomputing Center (BSC), Barcelona, Spain
45. Department of Biochemistry and Molecular Biomedicine, University of Barcelona, Barcelona, Spain
46. Pharmaceutical Sciences, University of Geneva, Geneva, Switzerland
47. Institute of Pharmaceutical Sciences of Western Switzerland, Geneva, Switzerland
48. Chemistry Department, University College London, London, United Kingdom
49. Swiss Bioinformatics Institute, Geneva, Switzerland
50. Center for Bioinformatics and Integrative Biology (CBIB), Facultad Ciencias de la Vida, Universidad Andrés Bello, Santiago, Chile
51. Department of Theoretical and Computational Biophysics, Max-Planck-Institute for Multidisciplinary Sciences, Göttingen, Germany
52. Department of Chemistry and Biochemistry, University of Oregon, Eugene, Oregon, United States
53. Department of Theoretical Condensed Matter Physics, Universidad Autónoma de Madrid, Madrid, Spain
54. School of Physics and Astronomy, University of Sheffield, Sheffield, United Kingdom
55. Pitzer Theory Center and Departments of Chemistry, Bioengineering, Chemical and Biomolecular Engineering, University of California, Berkeley, CA, United States
56. Department of Chemistry, Johns Hopkins University, Baltimore, Maryland, United States
57. Spanish National Institute of Bioinformatics (INB)/ELIXIR-ES, Barcelona, Spain
58. National Institute of Biological Sciences, Beijing, China
59. Department of Chemistry, Data Science Institute, University of Wisconsin-Madison, Madison, Wisconsin, United States
60. Department of Theoretical Biophysics, Max Planck Institute of Biophysics, Frankfurt am Main, Germany
61. Institute for Biophysics, Goethe University Frankfurt, Frankfurt am Main, Germany
62. NBD | Nostrum Biodiscovery, Barcelona, Spain
63. Department of Chemistry, University of Copenhagen, Denmark
64. Department of Electrical and Computer Engineering, Rutgers University, New Brunswick, New Jersey, United States
65. Ferrier Research Institute, Victoria University of Wellington, Wellington, New Zealand
66. Department of Chemistry, Yale University, New Haven, Connecticut, United States
67. Department of Chemistry – BMC, Uppsala University, Uppsala, Sweden
68. School of Chemistry and Biochemistry, Georgia Institute of Technology, Atlanta, GA, USA
69. School of Pharmacy and Centre for Biomolecular Sciences, University of Nottingham, Nottingham, United Kingdom
70. Department of Structural Biology, Stanford University School of Medicine, Stanford, California, United States
71. Faculty of Biomedical Sciences, Euler Institute, Università della Svizzera italiana (USI), Lugano, Switzerland
72. Department of Biochemistry and Biophysics, Science for Life Laboratory, Stockholm University, Solna, Sweden
73. Structural Biology and NMR Laboratory & the Linderstrøm-Lang Centre for Protein Science, Department of Biology, University of Copenhagen, Copenhagen, Denmark
74. Department of Chemistry, College of Staten Island, The City University of New York, New York, New York, United States
75. College of Pharmacy, Western University of Health Sciences, Pomona, California, United States
76. Institute of Biomedicine, University of Barcelona, Santa Coloma de Gramanet, Spain
77. Institute of Theoretical and Computational Chemistry, University of Barcelona, Santa Coloma de Garamet, Spain
78. Department of Pharmaceutical Sciences, School of Pharmacy, University of Maryland, Baltimore, Maryland, USA
79. National Research Council-Institute of Material Foundry, Scuola Internazionale Superiore di Studi Avanzati (SISSA), Trieste, Italy
80. Groningen Biomolecular Sciences and Biothechnology Institute, University of Groningen, Groningen, The Netherlands
81. Department of Chemistry and Biochemistry, University of California, San Diego, California, United States
82. Department of Pharmacology, University of California, San Diego, California, United States
83. Lerner Research Institute Cleveland Clinic, Cleveland, Ohio, United States
84. Department of Chemistry, Michigan State University, Michigan, United States
85. BioComp Group, Institute of Advanced Materials (INAM), Universitat Jaume I, Castellón, Spain
86. Center for Computational Chemistry, School of Chemistry, University of Bristol, Bristol, United Kingdom
87. Department of Chemical and Biological Engineering, Illinois Institute of Technology, Chicago, Illinois, United States
88. Department of Biotechnology, Bhupat and Jyoti Mehta School of Biosciences, Indian Institute of Technology Madras, Chennai, India
89. Department of Biomedical and Chemical Engineering, Syracuse University, Syracuse, New York, United States
90. Department of Physics, Freie Universität Berlin, Berlin, Germany
91. Department of Mathematics and Computer Science, Freie Universität Berlin, Berlin, Germany
92. Department of Chemistry, Rice University, Houston, Texas, United States
93. Microsoft Research AI4Science, Berlin, Germany
94. Department of Physics, University of York, York, United Kingdom
95. Department of Inorganic and Analytical Chemistry, Budapest University of Technology and Economics, Budapest, Hungary
96. Australian Institute for Bioengineering and Nanotechnology, The University of Queensland, Brisbane, Queensland, Australia
97. Department of Chemistry and Chemical Biology, Northeastern University, Boston, Massachusetts, United States
98. Center for Theoretical Biological Physics, Rice University, Houston, Texas, United States
99. Department of Physics and Astronomy, Rice University, Houston, Texas, United States
100. Department of BioSciences, Rice University, Houston, Texas, United States
101. Department of Computer Science, Virginia Polytechnic Institute and State University, Blacksburg, Virginia, United States
102. Department of Physics, Virginia Polytechnic Institute and State University, Blacksburg, Virginia, United States

103. Center for Soft Matter and Biological Physics, Virginia Polytechnic Institute and State University, Blacksburg, Virginia, United States
104. Institut de Química Computacional i Catàlisi (IQCC) and Departament de Química, Universitat de Girona, Girona, Spain
105. Catalan Institution for Research and Advanced Studies (ICREA), Barcelona, Spain
106. Department of Bioengineering, University of California Riverside, Riverside, California, United States
107. Department of Pathology and Molecular Medicine, School of Medicine, Queen's University, Kingston, Ontario, Canada
108. Department of Biology and Molecular Sciences, Queen's University, Kingston, Ontario, Canada
109. School of Computing, Queen's University, Kingston, Ontario, Canada
110. Biomolecular Simulations Group, Institut Pasteur de Montevideo, Montevideo, Uruguay
111. Department of Chemistry, Gottwald Center for the Sciences, University of Richmond, Richmond, Virginia, United States
112. Atomistic Simulations, Italian Institute of Technology, Genova, Italy
113. Department of Chemistry and Quantum Theory Project, University of Florida, Gainesville, FL, United States
114. Computational Biology Lab, Fundación Ciencia & Vida, Santiago, Chile
115. Facultad de Ingeniería y Tecnología, Universidad San Sebastián, Santiago, Chile
116. Department of Chemistry and Biochemistry, University of Delaware, Newark, Delaware, United States
117. Department of Physics & Beckman Institute, University of Illinois at Urbana-Champaign, Urbana, Illinois, United States
118. University of Texas Medical Branch, Galveston, Texas
119. Dipartimento di Scienze della Salute, Università di Catanzaro, Catanzaro, Italy
120. Laboratory of Theoretical Chemistry, Department of Chemistry, Sorbonne University, Paris, France
121. Department of Biosystems and Soft Matter, Institute of Fundamental Technological Research, Polish Academy of Sciences, Warsaw, Poland
122. Laboratory for Molecular Modeling, National Institute of Chemistry, Ljubljana, Slovenia
123. Department of Physics, Faculty of Mathematics and Physics, University of Ljubljana, Ljubljana, Slovenia
124. Department of Chemistry and Biochemistry, Faculty of Sciences, University of Porto, Porto, Portugal
125. Department of Biomedical Engineering, The University of Texas at Austin, Austin, Texas, United States
126. Department of Chemistry, University of Bergen, Bergen, Norway
127. Department of Physics and Astronomy, University College London, London, United Kingdom
128. Departament de Química Inorgànica i Orgànica (Secció de Química Orgànica) and Institut de Química Teòrica i Computacional (IQTCUB), Universitat de Barcelona, Barcelona, Spain
129. Department of Chemistry, University of Chicago, Chicago, Illinois, United States
130. Laboratory of Computational Chemistry and Biochemistry, Institute of Chemical Sciences and Engineering, Swiss Federal Institute of Technology (EPFL), Lausanne, Switzerland
131. Theoretical Biology and Biophysics, Los Alamos National Laboratory, Los Alamos, New Mexico, United States
132. New Mexico Consortium, Los Alamos, New Mexico, United States
133. Department of Chemistry and Courant Institute of Mathematical Sciences, New York University, New York, New York, United States
134. Department of Biology, Lomonosov Moscow State University, Moscow, Russia
135. Department of Chemistry, Stony Brook University, Stony Brook, New York, United States
136. Department of Applied Mathematics and Statistics, Stony Brook University, Stony Brook, New York United States
137. Biosciences Division and Center for Molecular Biophysics, Oak Ridge National Laboratory, Oak Ridge, Tennessee, United States
138. Department of Biochemistry and Cellular and Molecular Biology, University of Tennessee, Knoxville, Tennessee, United States
139. Computational Biophysics Research Team, RIKEN Center for Computational Science, Kobe, Japan
140. Theoretical Molecular Science Laboratory, RIKEN Cluster for Pioneering Research, Saitama, Japan
141. Laboratory for Biomolecular Function Simulation, RIKEN Center for Biosystems Dynamics Research, Kobe, Japan
142. Laboratory for Computational Molecular Design, RIKEN Center for Biosystems Dynamics Research, Kobe, Japan
143. Department of Chemistry, O'Donnell Data Science and Research Computing Institute, Center for Drug Discovery, Design, and Delivery (CD4), Southern Methodist University, Dallas, Texas, United States
144. School of Pharmacy, Queen's University Belfast, Belfast, United Kingdom
145. Departamento de Química Física, Universidad de Valencia, Burjassot, Spain
146. School of Biochemistry, University of Bristol, Bristol, United Kingdom
147. Department of Chemistry, Chicago Center for Theoretical Chemistry, Institute for Biophysical Dynamics, and James Franck Institute, The University of Chicago, Chicago, IL, United States
148. Molecular and Cellular Modeling Group, Heidelberg Institute for Theoretical Studies (HITS), Heidelberg, Germany
149. Department of Chemistry, University of Southern California, Los Angeles, California, United States
150. Macromolecules Innovation Institute, Virginia Tech, Blacksburg, Virginia, United States
151. Department of Chemistry and Biochemistry, University of Lethbridge, Lethbridge, Alberta, Canada
152. Department of Chemistry and Biochemistry, University of Missouri-St. Louis, St. Louis, Missouri, United States
153. Institute of Bioinformatics and Structural Biology, National Tsing Hua University, Hsinchu, Taiwan
154. Physics Department, Technical University of Munich, Garching, Germany

The communities that embraced data archiving efforts decades ago are now, in the era of data-driven biology, those gaining the most from the AI revolution. The structural biology community was a pioneer in this regard by establishing the Protein Data Bank (PDB) in 1971 and making data accessible using the FAIR (Findable, Accessible, Interoperable and Reusable) principles even before these were articulated [1, 2]. The genomics/bioinformatics community has followed the example, establishing many widely used databases [3, 4]. By contrast, molecular simulation has been anchored in usage paradigms dating back to the seventies, when molecular dynamics (MD) simulation was first applied to study biomacromolecules [5]. At that time, MD was used by theoretical physicists and chemists in "proof of concept" simulations. Still, fifty years later, MD has evolved into a cornerstone molecular biology technique that can provide accurate quantitative analysis and property prediction. MD is now employed by tens of thousands of researchers worldwide, accounting for roughly 15% of global supercomputer usage. Unfortunately, these rich and costly data are not systematically maintained, and when further analyses are required, simulations have to be rerun—an unacceptable situation from scientific, environmental and sustainability standpoints. In this letter, we argue for a collaborative endeavor to archive MD simulation data and describe ongoing efforts to establish cost-effective and sustainable data archiving strategies.

Advances in computer technology have made it possible to simulate large realistic biological systems beyond the millisecond timescale, and we are seeing simulations in the $10^9$ particle range, covering entire organelles and even minimal cells, resulting in a "deluge of data" [6] in a field that lacks agreed strategies for data storage. As in the seventies of the last century, trajectories obtained after a huge effort are often ignored (or even deleted) after a hypothesis-driven analysis presented in a scientific publication. For a field entirely based on *sampling,* and where the recipe for observations can be described exactly and critically assessed, this is a huge problem! Instead of being able to re-analyse, reuse, and potentially spot undetected artifacts or new features in data, readers are often expected to blindly trust the closed set of statements made by the authors in a paper. The lack of a systematic approach to store data (and associated metadata) prevents new studies based on previous trajectories, impedes meta-analyses, the training of machine-learning approaches, optimization of force fields and simulation protocols, generation of new conformations for modeling of reactivity, hampers the use of trajectories to train coarse-grained and mesoscopic models, or generative models, and prohibits the integration of MD-results into the rich ecosystem of bio-databases. Some journals have started requiring the repository of trajectories, further indicating that the community needs to escape from a paradigm that made sense in the seventies, but that now hinders progress, and move to an open-science model.

Establishing an archive for biosimulation data - upon quality assessment - would address these issues, democratize the field, and materialize the impact of MD simulations on life-science research. The traditional view held by the simulation community that storing and archiving is more expensive than re-computing, which might have been correct in the past, is no longer valid, as demonstrated by the massive Folding@home study on the SARS-CoV-2 main protease [7], or for many millions of atom simulations [8, 9]. However, the new science that can be learned from stored trajectories is more important than the cost. For instance, the ABC Consortium [10] was established in 2004 as a community effort generating a Gigabytes-database of DNA simulations, which had grown to hold 15 Terabytes of data by 2019. The original goal of ABC was to study DNA polymorphisms, but the database has become crucial in other fields, like the study of signal transfer in DNA and the development of coarse-grained models. The current HexABC database contains 400 Terabytes of data generated to explore 6-mer dependencies of DNA dynamics. However, its future use, which is currently difficult to anticipate, might be more important than the current goals of the project. Another wonderful example emerged during the COVID-19 pandemic [11, 12, 13], when the Molecular Sciences Software Institute (MolSSI) [14], in collaboration with European groups including BioExcel, EOSC, EBI, and Zenodo created the COVID-19 Molecular Structure and Therapeutics Hub that went live in April 2020, connecting scientists across the global biomolecular simulation community, as well as improving the connection between simulation and experimental and clinical data and their investigators (covid.molssi.org). A further example is MDverse, an effort to make MD trajectories FAIRer by indexing and curating thousands of simulations scattered across the internet [15].

The challenges that lie ahead for the community are diverse. The technical ones, namely sustained data storage capacity, bandwidth, and processing capacity for analysis, can be alleviated by a distributed database policy following initiatives such as the EGA infrastructure (European Genome-Phenome Archive; [16]) and by the commitment of funding institutions and HPC centers, offering storage, bandwidth, and processing capabilities. A centralized management entity should coordinate the federated nodes, defining required metadata, deposition policies, guaranteeing compliance of FAIR rules and providing a common entry point through web-based and programmatic REST API interfaces. The myriad of variants of MD programs, protocols, formats, and simulation conditions used lead to more complex problems [17]. Recent MD repositories and databases [13] are already prepared to manage not only plain MD trajectories but also Markov State Models, ensembles, multiscale simulations (hybrid or combined mesoscale/CG/atomistic, QM/MM), constant pH, replica exchange, and MD trajectories biased with metadynamics or similar methods. NoSQL databases such as MongoDF/GridFS allow efficient storing and querying of the diversity of outputs provided by MD engines and are already adopted by MD storage initiatives. However, much more work is required for an effective analysis framework that can manage an increasingly larger number of MD variants and trajectory formats.

Data should be findable, with each entry registered with a persistent identifier, ideally a DOI, ensuring a proper citation, following the example of the WorkflowHub registry (https://workflowhub.eu/; [18]). Furthermore, they should be stored in an interoperable manner, so that it can be read and exploited by current and future data scraping and Machine Learning (ML) algorithms. To this end, the community must reach an agreement to standardize MD data exchange formats with: i) efficient trajectory compression including simple system specifications (e.g. atom/residue names and connectivity); ii) key-value trees storing high-level and full simulation settings metadata; and iii) metadata-based ontology [19, 20], which would allow the user to search databases based on the contents, the nature, or even the purpose of the simulations. Standardized provenance [21] should be stored by means of data blocks specifying commands/operations used to generate the trajectory together with names, stored hash sums of the complete files used for input, and specific software used (with precise versions). This would allow the user to reproduce all the different steps followed to prepare and run the simulation, including modeling of missing residues, physical conditions (e.g. pH, salt concentration, temperature and pressure), force fields, especially methodology used to obtain parameters involving non-standard molecules (e.g. small molecules, membrane systems, ionic coordination), as well as equilibration and possibly sampling process. Minimum metadata should include system information, simulation parameters, author(s), data license and copyright, and, importantly, the main purpose of the simulation. The definition of standardized protocols (i.e., list of operations) for production run and analysis, including a troubleshooting section could be added [22]. These, and a set of metadata dependent quality-control analysis, both general and system specific, are crucial requisites for gaining trust from the community and for defining deposition rules. A data repository following FAIR principles and the associated analysis tools will increase the impact and the reproducibility (complex at the binary level) of MD in related fields in the Life-Science Data ecosystem, from genomics to structural biology and from protein and drug design to molecular biology [23]. MD data would provide unique dynamic information of biological macromolecules fully complementary with the rich information available from PDB. This could be integrated into the Life Science ecosystem following the approach of the Protein Data Bank in Europe Knowledge Base (PDBe-KB), designed for the integration and enrichment of 3D-structure data and functional annotations [24]. All this information will contribute to the *knowledge democratization,* helping research teams with limited resources and fueling further advances in Artificial Intelligence (AI) in the scientific domain [25].

The MDDB project (https://mddbr.eu/) and similar initiatives aim to establish such a repository, allowing: i) data quality assessment metrics to increase the trust of the community in the deposited data; ii) common data format, metadata requirements, and ontologies to facilitate interoperability; iii) a minimum set of information needed to store and reproduce the simulations, including data provenance, license and copyright; and iv) a standard and robust infrastructure to store and share the data, with persistent identifiers and different ways to access them. We believe science will be better served by fully embracing this data-driven view of biomolecular simulation. Furthermore, data-driven initiatives such that supported by this letter, would help the interaction with other simulation communities, such as the material science one, which share some of the problems the biomolecular simulation community is facing.

Graphical Abstract:

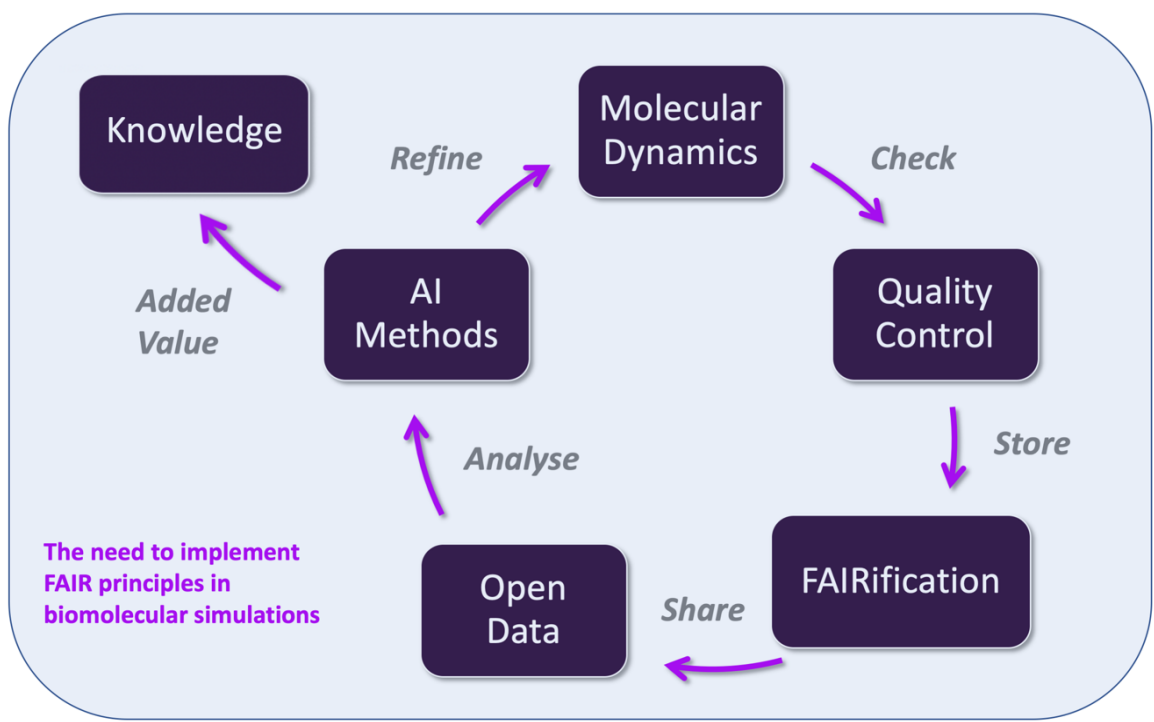